\documentclass{llncs}
\usepackage{listings}
\usepackage{epsfig}
\usepackage{makeidx}

\begin{document}

\frontmatter          % for the preliminaries

\pagestyle{headings}  % switches on printing of running heads

\mainmatter              % start of the contributions
\title{A Type-Oriented Graph500 Benchmark}
\titlerunning{A Type-Oriented Graph500 Benchmark}  % abbreviated title (for running head)
\author{Nick Brown}
\authorrunning{Nick Brown} 
\tocauthor{Nick Brown}
\institute{EPCC, Edinburgh University\\
\email{nick.brown@ed.ac.uk}}

\maketitle              % typeset the title of the contribution

\begin{abstract}
Data intensive workloads have become a popular use of HPC in recent years and the question of how data scientists, who might not be HPC experts, can effectively program these machines is important to address. Whilst using models such as Partitioned Global Address Space (PGAS) is attractive from a simplicity point of view, the abstractions that these impose upon the programmer can impact performance.
We propose an approach, type-oriented programming, where all aspects of parallelism are encoded via types and the type system which allows for the programmer to write simple PGAS data intensive HPC codes and then, if they so wish, tune the fundamental aspects by modifying type information.
This paper considers the suitability of using type-oriented programming, with the PGAS memory model, in data intensive workloads. We compare a type-oriented implementation of the Graph500 benchmark against MPI reference implementations both in terms of programmability and performance, and evaluate how orienting their parallel codes around types can assist in the data intensive HPC field.
\end{abstract}

\section{Introduction}
The HPC community has traditionally concentrated on solving computation based problems but in recent years data intensive workloads have also become a popular use of these resources. Data intensive workloads often involve huge amounts of data, each requiring small numbers of calculations per element and because of this the communication aspects of a system is critically important. This is in contrast with the more traditional computation based workloads, where data sizes tend to be smaller but much more computation per element is required. One of the challenges associated with HPC is  programming models and the data processing field is no exception. There is often a trade off between programmability and performance; those models which promote simplicity can impose choices and restrictions upon the programmer in the name of abstraction which can harm performance. Whilst efficient communication is important to both computation and data intensive workloads, the fact that data intensive workloads place so much emphasis on communication makes it critically important that the programming models used do not sacrifice communication efficiency.

Using the PGAS memory model for solving data intensive problems, where data is equally accessible whether it is held in local or remote memory and the programmer need not worry about the underlying implementation detail, is an attractive proposition from a programmability point of view. However, this higher level of abstraction can typically result in a performance cost and existing PGAS languages either disallow or limit the options that the programmer has to tune their code at the communication level. Type-oriented programming addresses the PGAS trade off issue by providing the options to the end programmer to choose between explicit and implicit parallelism by using types which can be combined to form the semantics of data, governing parallelism. A programmer may choose to use these types or may choose not to use them and in the absence of type information the compiler will use a well-documented set of default behaviours. Additional type information can be used by the programmer to tune or specialise many aspects of their code which guides the compiler to optimise and generate the required parallelism code. In short these types for parallelisation are issued by the programmer to instruct the compiler to perform the expected actions during compilation and in code generation.

The Graph500\cite{graph500} benchmark is a popular, objective, way of determining hardware's suitability to data processing but can also be used to benchmark other aspects such as the languages, libraries and runtimes that are used in data intensive work. We compare an implementation of the Graph500 benchmark in Mesham, our research type-oriented programming language, compared to the MPI reference versions both in terms of programmability and performance. 

\section{Background}
\subsection{Type-oriented programming}
Type-oriented programming\cite{types} allows the programmer to encode all variable information via the type system by combining different types together to form the overall meaning of variables. This is contrasted against a more traditional approach where the programmer uses a type to govern the data that a variable will hold but additional information, such as whether a variable is read only or not, is applied via type qualifiers. Using the C programming language as an example, in order to declare a variable m to be a read only character where memory is allocated externally, the programmer writes \emph{extern const char m}. Where \emph{char} is the type and both \emph{extern} and \emph{const} are inbuilt language type qualifiers. Whilst this approach works well for sequential languages, in the parallel programming domain there are potentially many more attributes which might need to be associated; such as where the data is located, how it is communicated and any restrictions placed upon this. Representing such a rich amount of information via multiple qualifiers would not only bloat the language, it might also introduce inconsistencies when qualifiers were used together with potentially conflicting behaviours.

Instead our approach is to allow for the programmer to combine different types together to form the overall meaning. For instance, \emph{extern const char m} becomes \emph{var m:Char::const::extern}, where \emph{var m} declares the variable, the operator \emph{:} specifies the type and the operator \emph{::} combines two types together. In this case, a \textbf{type chain} is formed by combining the types \emph{Char}, \emph{const} and \emph{extern}. Precedence is from right to left where, for example, the read only properties of the \emph{const} type override the default read \& write properties of \emph{Char}. It should be noted that some type coercions, such as \emph{Int::Char} are meaningless and so rules exist within each type to govern which combinations are allowed.

Within type-oriented programming the majority of the language complexity is removed from the core language and instead resides within the type system. The types themselves contain specific behaviour for different usages and situations. The programmer, by using and combining types, has a high degree of control which is relatively simple to express and modify. Additionally, by writing code in this high level way means that there is a rich amount of information upon which the compiler can use to optimise the code. In the absence of detailed type information the compiler can apply sensible, well documented, default behaviour and the programmer can further specialise this using additional types if required at a later date. The result is that programmers can get their code running and then further tune if needed by using additional types.

Mesham\cite{mesham} is a programming language that we have developed to research and evaluate the type-oriented paradigm. It follows a simple imperative language with extensions to support this type-oriented paradigm and a Partitioned Global Address Space memory model. Using the PGAS memory model, the entire global memory, which is accessible from every process, is partitioned and each block has an affinity with a distinct process. Reading from and writing to memory (either local or another processes' chunk) is achieved via normal variable access and assignment. The benefit of PGAS is its conceptual simplicity, where the programmer need not worry about the lower level and often tricky details of communication. This makes it an ideal memory model for data scientists, who are experts in using their own data but not necessarily HPC.

Type-oriented programming provides the best of both worlds. In Mesham by default all communication is one sided however this can be overridden using additional type information which further tunes and specialises the communication behaviour of specific variables. The programmer can get their codes working and then tune, by using types, fundamental aspects such as communication to improve performance and/or scalability. This is contrasted against traditional HPC programming models, where fundamental aspects are often integral to the code and changing them can require widespread changes or even a code rewrite.

\subsection{Graph 500 benchmark}
As data intensive workloads are fundamentally limited by communication, existing computation based benchmarks such as LINPACK are of limited use. Instead, the Graph500 benchmark\cite{graph500} has been developed to stress the communication aspects of a system. It consists of three phases; graph construction which constructs graph in Compressed Sparse Row (CSR) format, a Breadth First Search (BFS) kernel and lastly the a validation of the BFS traversal. The Graph500 problem size is represented using scale and edge factors. The scale is the logarithm base two of the number of vertices and edge-factor is the ratio of the graph’s edge count to its vertex count. Therefore a graph has 2\textsuperscript{scale} vertices and 2\textsuperscript{scale}$\times$edge-factor edges. Performance of the BFS is measured in Traversed Edges Per Second (TEPS.)
A number of reference implementations are provided and of these there are four MPI based codes, a one-sided implementation where vertex communications are done using MPI-2 one-sided communications, an MPI simple version which uses asynchronous point to point communications,
an MPI replicated compressed sparse row implementation and a MPI replicated compressed sparse column code. Whilst all of these implementations use the level synchronized BFS traversal algorithm \cite{BFS} their implementations all require separate codes and to go from one to another has required substantial code rewriting of the BFS kernels. In this paper we concentrate on the one-sided and the simple (asynchronous point to point) versions which, at an algorithm level, only differ in terms of their form of communication but require entirely separate kernel implementations. 
In reality data scientists do not want to be working at this lower level; they want to be able to write a simple code and then easily modify fundamental aspects, such as the form of communication, to experiment with parallelism.

\subsection{Related work}
High Performance Fortran(HPF)\cite{hpf} is a parallel extension to Fortran90. The programmer specifies just the data partitioning and allocation, with the compiler responsible for the placement of computation and communication. The type-oriented approach differs because programmer can, via types, control far more aspects of parallelism. Alternatively, if not provided, the type system allows for a number of defaults to be used instead. Co-array Fortran (CAF)\cite{caf} provides the programmer with a greater degree of control than in HPF, but still the method of communication is implicit and determined by the compiler whilst synchronisations are explicit. CAF uses syntactic shorthand communication commands and synchronisation statements hard wiring these into a language is less flexible than our use of types. Chapel\cite{chapel} is another PGAS language and supports the programmer controlling aspects of parallelism by providing higher and lower levels of abstractions. Many of these higher level constructs in Chapel, such as reduction are implemented via inbuilt operators and keywords, contrasted to Mesham where they would be types in an independent library.

Co-array C++ \cite{coarrayc++} integrates co-arrays into C++ using template libraries. The C++ programmer adds additional information to their source code through these template libraries which determine parallelism. Whilst the type library of Mesham has a far wider scope than the current co-array template library it would be possible to encode our types as a C++ template library. This illustrates how the core language itself is actually irrelevant and our approach could be applied to existing languages, such as C++. The benefit of this is that programmers could orient their parallelism around types within a familiar language. The downside of this approach is that the actual C++ language is fixed and whilst template libraries are well integrated, the flexibility of our use of types would be limited to the current C++ approach and compile time optimisation might be limited.

Parallelizing ARRAYs (PARRAY)\cite{parray} extends the C and C++ languages with new typed arrays that contain additional information such as the memory type, layout of data and the distribution over multiple memory devices. The central idea is that a programmer need only learn one unified style of programming and this applies equally to all major parallel architectures. The compiler will generate code according to the typing information contained in the source. Whilst this approach is similar to the types used in Mesham, there are some important differences. As a bolt on to existing languages, PARRAY uses its own syntax, similar to pre-processor directives, to declare arrays with types; for instance \emph{pinned}, \emph{paged} or \emph{dmem} are used denote which device holds the array. In dealing with the arrays PARRAY still requires a number of inbuilt commands to handle its data structures. Mesham takes this a stage further and types are fully integrated into the language which means that, instead of requiring commands to copy or transpose data, language operators such as assignment will automatically handle the operation according to the type information. This integration is central to the data intensive example considered here where parallelism is entirely integrated in the language; for instance references to data may be local or global but in the simplest case the programmer need not worry about distinguishing these aspects to get their code working. In our approach there many types which the programmer can use to tune their data structures but equally omitting these is fine which will result in some safe default behaviour being applied.

The approach that PARRAY follows, bolting on parallelism using some syntax to differentiate it from the existing language is familiar. Solutions such as OpenMP\cite{openmp} allow for the programmer to direct parallelism through pre-processor directives which guide the compiler how to handle parallelism. Importantly in our approach, types are first class citizens in the language so integrate fully with the existing language semantics which means that the programmer has the flexibility to support aspects such as creating new types in their code and reasoning about type information using existing language constructs. Through constructing type chains we provide a mechanism for building up complex type information in a structured manner and it is this type chain that provides the semantics of operations performed on the variable.

\section{Implementation}
\begin{lstlisting}[frame=lines,caption={Default one sided communication},label={lst:bfsonesided}]
typevar GraphVertex::=referencerecord["children", array[GraphVertex],"numChildren", Long, "id", Long];
var vertices:array[GraphVertex,nvtx_scale] :: allocated[partitioned[numProcs] :: single[evendist]]

var searchTree:array[Long, nvtx_scale] :: allocated[partitioned[numProcs] :: single[evendist]];
var childrenParents:array[Long, nvtx_scale] :: allocated[partitioned[numProcs] :: single[evendist]];
var globalNextVerticies:=1;
var vertexQueue:queue[GraphVertex] :: allocated[multiple];
var vertexQueueNext:queue[GraphVertex] :: allocated[multiple];

if (vertices[rootVertexIndex].on == myPid) { vertexQueue:=root; childrenParents[root.id]:=root.id; };
while (globalNextVerticies > 0) {
  while (!vertexQueue.empty) {
    var singleVertex:GraphVertex;
    singleVertex:=vertexQueue;
	if (searchTree[singleVertex.id] == -1) {
	  searchTree[singleVertex.id]:=childrenParents[singleVertex.id];
	  for i from 0 to singleVertex.numChildren - 1 {
		var childVertex:=singleVertex.children[i];
		childrenParents[childVertex.id]:=singleVertex.id;	
		vertexQueueNext.on[childVertex.on]:=childVertex;
  }; }; };
  vertexQueue:=vertexQueueNext;
  vertexQueueNext.clear;
  globalNextVerticies::allreduce["sum"]:=vertexQueue.size; };
\end{lstlisting}
Listing \ref{lst:bfsonesided} illustrates the Mesham source code for the BFS kernel and associated variable declarations. For clarity we concentrate on the core BFS kernel, hence supporting functions such as building the Kronecker graph of vertices, constructing the edge list, finding search keys and validating the resulting search tree have been omitted from this paper. The \emph{typevar} keyword at line 1 creates a new type which is called a \emph{GraphVertex} and is basically an alias for the \emph{referencerecord} type which is a record and similar to a struct in C. This record is used to represent a graph vertex and the \emph{referencerecord} type supports members of a record referencing other records. An example of this is the array at line 1, \emph{children}, which contains references to other vertex records which are the children of this vertex. The references themselves can either be to local data or point to global data which is held on another process. In terms of correctness, the programmer need not distinguish between local and global data references as the type library takes care of the underlying communications required; although they might want to differentiate for performance reasons.

Lines 2, 4 and 5 set up arrays \emph{vertices}, the actual graph vertices which have already been built up, \emph{searchTree}, the search tree to return from this kernel holding the resulting vertex parent ids and \emph{childrenParents} which holds the parent ids of children vertices for the next level. Using the \emph{allocated} type the programmer has provided some additional information to guide the compiler in how to allocate this data. Combined with the \emph{single} type it means that a single copy of the array exists globally, which is split up into \emph{numProcs} partitions which are then evenly distributed amongst the distinct process memories via the \emph{evendist} type. Because these three arrays are the same size they are distributed in the same manner with the same indexes (and hence vertices) on each process.

Lines 7 and 8 create two queues, one for the current BFS level and one for the next BFS level with each queue holding data of type \emph{GraphVertex}. The use of the \emph{multiple} type as an argument to the \emph{allocated} type informs the compiler that each process will hold a distinct version of these variables in their own memory. Each type determines the behaviour whenever a variable is used, an example of this is at line 10 where the \emph{referencerecord} type implements the \emph{on} operator which returns the process which holds the actual data which the reference is pointing to. In this case the effect is to add the root search vertex to the current level queue on that specific owning process. The effect of the assignment at line 14 will be to pop the top most vertex from the local queue and place it into the \emph{singleVertex} variable, the fact that this is a pop is because of the appropriate types of the variables involved in the assignment and for the same reason a queue addition is done at lines 20 and 10 but line 22 copies the entire \emph{vertexQueueNext} into \emph{vertexQueue}.
For each level, a process will iterate through their vertex queue. For each vertex if it has not already been processed (line 15) then the process will iterate through each child. Each child is added to the next level vertex queue on the appropriate holding process as well as that child's parent id to the \emph{childrenParents} array. In the absence of further type information, how these communications occur is entirely abstracted away and if the child is on a different process to the parent then the default behaviour will be one sided communication. Part of a global reference is which process's memory actually holds the data. This means that a \emph{referencerecord}'s \emph{.on} operation is a local operation on the reference itself and the vertex communication required in this BFS implementation is to place remote vertices on their own processes' next level queue; much of this remote placing can be implemented with minimal communication. 
At line 24 there is a global all-reduce to determine the number of vertices to be processed at the next level and the algorithm will terminate if, at the global level, there are none. This \emph{globalNextVerticies::allreduce["sum"]} illustrates an additional aspect of type-oriented programming where the type behaviour of a variable can be overridden by the additional of extra types for a specific expression; for this assignment only a blocking all-reduce is issued.

It can be seen that this is a simple, high level algorithm with the underlying types taking care of much of the lower level and tricky implementation details. Compared to the existing one sided MPI reference code this implementation is far shorter, 28 lines of code compared to 205 in the reference implementation and allows the programmer to concentrate on the algorithmic and data structures of their code. It is true that there are underlying Mesham types and runtime libraries to support the code which amount to 500 lines of code, however, these are very general an can be reused in multiple codes; the partition and distribution types we used for this benchmark were originally written for a Mesham asynchronous Jacobi implementation \cite{meshameasc}. It illustrates how an HPC expert can construct these types once and these then can be used time and time again in different contexts. 
By simplifying the code, a data scientist, who might not be an HPC expert, will be able to get their code working and to a reasonable performance level. Whilst the default one sided communication is a simple and safe behaviour it is often not particularly efficient. As already mentioned, one of the MPI reference implementations is a point to point code, which replaces the one sided communication with asynchronous point to point and greatly improves the efficiency. However, to achieve this the core BFS code has had to be rewritten and additional low level issues such as matching asynchronous communications and buffer sizes which are complex and error prone has had to be considered. By orienting all aspects of parallelism around types the programmer can get their code working and then further tune for performance and listing \ref{lst:bfsp2p} sketches the modified Mesham code of listing \ref{lst:bfsonesided} to use asynchronous point to point communication rather than the default one-sided.
\begin{lstlisting}[frame=lines,caption={Asynchronous p2p communications},label={lst:bfsp2p},numbers=none]
var childrenParents:array[Long, nvtx_scale] :: allocated[partitioned[numProcs] :: single[evendist]] :: async;
var vertexQueue:queue[GraphVertex] :: allocated[multiple] :: async;
var vertexQueueNext:queue[GraphVertex] :: allocated[multiple] :: async;
...
while (globalNextVerticies > 0) {
  while (!vertexQueue.empty) {
    ...};
  sync;
  vertexQueue:=vertexQueueNext;
  ... };
\end{lstlisting}
The code structure has remained the same and minimal changes, mainly oriented around the types, have been made to modify the underlying communication method. The addition of the \emph{async} type to the \emph{childrenParents} variable and queues guide the compiler to use asynchronous point to point communication for all communications involving these variables. The assignment at line 19 (listing \ref{lst:bfsonesided}) now issues an asynchronous send of the parent vertex id and at line 20 an asynchronous send as part of the remote queue addition. The \emph{sync} keyword has been added at the end of processing the current level in listing \ref{lst:bfsp2p} and waits for all outstanding asynchronous communications to complete, ready to proceed with the next search level. Effectively the addition of these \emph{async} types set up the same asynchronous message listening and coalescing buffers that are present in the reference MPI implementation but these low level details are abstracted away from the programmer. In the absence of further type information the size of the coalescing buffer is set to be 256 \emph{GraphVertex} elements, although this can be further tuned by providing an argument to the \emph{async} type such as \emph{async[128]}. This is an example of where the meaning of the argument provided to types depends entirely on the type chain and it is within the context of, in this case, the \emph{queue} and \emph{async} types to interpret the arguments accordingly.
\section{Results}
\begin{center} 
\begin{figure}[htb]
\begin{center}\includegraphics[scale=0.5]{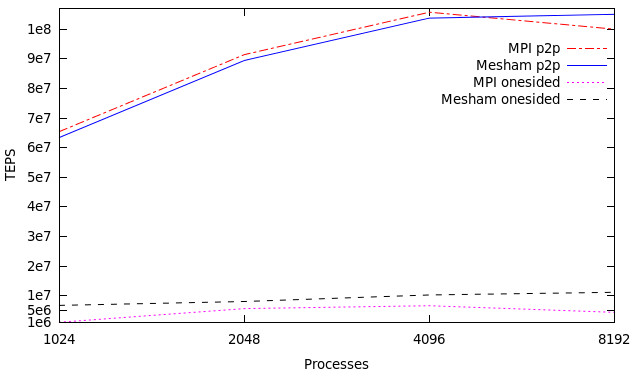}\end{center}
 \caption{Performance of Mesham vs MPI benchmarks}
    \label{fig:performancecombined}
\end{figure}
\end{center}
Figure \ref{fig:performancecombined} illustrates the strong scaling characteristics of our Mesham and MPI BFS implementations on a Cray XE6 using a vertex scale of 29. The upper plots, labelled \emph{p2p}, are the asynchronous Mesham point to point BFS implementation and the reference MPI simple implementation. It can be seen that the performance of the Mesham and MPI versions are comparable, with the Mesham version slightly under performing the MPI implementation but the difference is small. Initially in both versions the TEPS increases as the number of cores is increased but, commonly with strong scaling experiments, a point is reached where the cost of communication outweighs the benefits gained from additional parallelism and performance starts to degrade. In the results we can see that the MPI implementation actually performs worse over 8192 cores than on 4096 cores, and the Mesham version's TEPS at 8192 cores is only a slight improvement over the 4096 core run.

Figure \ref{fig:performancecombined} also depicts the strong scaling performance of the default, one sided, Mesham implementation and the one sided MPI benchmark which are the lower two plots and labeled \emph{onesided}. This illustrates that the safe and simple behaviour incurs a performance hit which can then be tuned using additional types to the performance in the p2p case. The Mesham one sided implementation out performs the MPI one-sided implementation due to the compiler optimising communications and one sided epoch windows which is possible because of the rich amount of type information available. This illustrates, in itself, a performance benefit of writing high level data intensive codes using Mesham default communications compared to a lower level implementation.

\section{Conclusions and further work}
In this paper we have considered how type-oriented programming may be applied to the data intensive HPC field. Aspects of this paradigm could, in the future, be used as part of existing languages to achieve the best of both worlds; the advantages discussed in this paper along with the familiarity of existing models and languages. 
We have shown that, by using types, the programmer can write conceptually simple PGAS style data processing codes at no significant hit in performance compared to traditional implementations. Types provide the additional benefit that the programmer can initially concentrate on the correctness of their codes and then, once a simple working version exists, they can use types to tune for performance as illustrated in the the Mesham one sided and point to point BFS implementations.
There is further work to be done understanding the reasons behind the slight performance gap of the Mesham and MPI implementations and based upon this work we are now looking to examples of data intensive problems, rather than benchmarks, to understand how this paradigm can help in solving real world data intensive workloads.


\begin{thebibliography}{}
\bibitem{graph500}
Murphy, B., Wheeler, K., Barrett, B., Ang, J.:
Introducing the Graph 500.
In: Cray User's Group (CUG), 2010
\bibitem{types}
Brown, N.:
Applying Type Oriented Programming to the PGAS Memory Model.
In: 7th International Conference on PGAS Programming Models (PGAS), 2013
\bibitem{mesham}
Brown, N.:
The Mesham language specification.
http://www.mesham.com , 2013
\bibitem{BFS}
Suzumura, T., Ueno, K., Sato, H., Fujisawa, K., Matsuoka, S.: 
Performance Characteristics of Graph500 on Large-scale Distributed Environment. 
In: 2011 IEEE International Symposium on Workload Characterization, IISWC 2011
\bibitem{hpf}
Luecke, G., Coyle, J.:
High Performance Fortran Versus Explicit Message Passing On The ISB SP-2
In: Technical Report Iowa State University, 1997
\bibitem{caf}
Numrich, R., Reid, J.:
Co-array Fortran for parallel programming
In: ACM SIGPLAN Fortran Forum, 1998
\bibitem{coarrayc++}
Johnson, T.:
Coarray C++
In: 7th International Conference on PGAS Programming Models (PGAS), 2013
\bibitem{chapel}
Cray Inc. Seattle:
Chapel language specification (version 0.82)
http://chapel.cray.com/ , 2011
\bibitem{parray}
Chen, Y., Xiang, C. Hong, M.:
PARRAY: A Unifying Array Representation for Heterogeneous Parallelism
In: Proceedings of the 17th ACM SIGPLAN Symposium on Principles and Practice of Parallel Programming, 2012
\bibitem{openmp}
OpenMP Architecture Review Board:
OpenMP Application Program Interface Version 3.0
http://www.openmp.org/mp-documents/spec30.pdf , 2008
\bibitem{meshameasc}
Brown, N.:
Type oriented programming for exascale.
In: Exascale Applications and Software Conference (EASC), 2013
\end{thebibliography}
\end{document}